\documentclass[aps,pra,twocolumn,epsfig,showpacs]{revtex4}

\usepackage{dcolumn}
\usepackage{bm}
\usepackage{graphicx}
\usepackage{amsmath}
\usepackage{latexsym}
\usepackage{amsfonts}
\usepackage{amssymb}
\usepackage{array}
\usepackage{epsfig}

\newcommand{\ket}[1]{\left\vert#1\right\rangle}
\newcommand{\bra}[1]{\left\langle#1\right\vert}

\begin{document}

\title{Effects of squeezing on quantum nonlocality of superpositions of coherent states }
\author{Chang-Woo Lee and Hyunseok Jeong}
\affiliation{
Center for Theoretical Physics,
Center for Subwavelength Optics and Department of Physics and Astronomy,
Seoul National University, Seoul, 151-742, Korea}
\date{\today}

\begin{abstract}
We analyze effects of squeezing upon superpositions of coherent states (SCSs) and entangled coherent states (ECSs) for Bell-inequality tests. We find that external squeezing can always increase the degrees of Bell violations, if the squeezing direction is properly chosen, for the case of photon parity measurements. On the other hand, when photon on/off measurements are used, the squeezing operation can enhance the degree of Bell violations only for moderate values of amplitudes and squeezing. We point out that a significant improvement is required over currently available squeezed SCSs in order to directly demonstrate a Bell-inequality violation in a real experiment.
\end{abstract}

\pacs{03.65.Ud, 42.50.Dv, 42.50.-p}
\maketitle

\section{Introduction}
Einstein, Podolsky, and Rosen (EPR)
questioned completeness of quantum mechanics based on the idea of local realism \cite{EPR}.
Bell suggested a profound and useful inequality imposed by local hidden variable theories,
which reflects EPR's idea \cite{Bell}. A couple of refined versions of Bell's inequality
followed the original one \cite{CHSH,CH}, and numerous experimental demonstrations have
also been performed \cite{Freedman, Aspect}. In these studies, quantum states of light
have played a crucial role. Indeed, all Bell inequality tests in which
the space-like separation between two local parties is satisfied have been performed using photons.
In the meantime, it is worth noting that a loophole-free Bell inequality test is yet to be performed.
The major obstacle in typical photon-based experiments, where two local parties
are separate enough, is probably the detection loophole \cite{Garg87}.
Very recently, a Bell inequality test free from the detection loophole was performed
using remote atomic qubits \cite{Mat08}, however, it did not satisfy
the space-like separation required for a loophole-free Bell test.

Recently, various types of continuous-variable states have been studied in order to suggest
proposals for loophole-free Bell inequality tests \cite{Nhaetal}.
As non-Gaussian continuous-variable states have rich structures in the phase space,
it is important to explore possibility of efficient Bell inequality tests using those states.
Among non-Gaussian continuous-variable states,
superpositions of two coherent states (SCSs) \cite{SCS,WScat}
in free-traveling optical fields have been found a very useful tool
for fundamental tests of quantum theory \cite{Derek,Jeong01,Magda,An,psss08,realism09}
as well as for quantum information applications
 \cite{Enk01,JKL01,Jeong02,Ralph03,WeakForce,puri}.
In particular, they are useful for Bell inequality tests
using various measurements such as photon on/off detection,
photon number detection, and homodyne detection \cite{Derek,Jeong01,Magda,An}.
Once single-mode SCSs are generated,
a 50:50 beam splitter can be used to
generate entangled coherent states (ECSs) \cite{ECS}
with which one can perform Bell-inequality tests \cite{Derek,Jeong01,Magda,psss08,realism09}.

Recently, ``squeezed'' SCSs were generated and detected
\cite{Ourjoumtsev,Sasaki1,Sasaki2}, where the size of the states ($\alpha=\sqrt{2.6}$) was
reasonably large for fundamental tests of quantum theory and
implementations of quantum information processing \cite{lund08}.
Squeezed SCSs can be more robust against decoherence than unsqueezed ones \cite{Serafini}
while they have similar nonclassical properties and usefulness in quantum information applications
\cite{Obada,suitability,Dell}.
Remarkably, it has been clearly pointed out that the squeezed SCSs recently generated
can be used for proof-of-principle experiments such as quantum teleportation
and single qubit gates without any modifications \cite{suitability}.
This strongly motivates us to study effects of squeezing on
SCSs and ECSs for various purposes.

In this paper, we study effects of squeezing on SCSs and ECSs for the purpose of
Bell-inequality tests using
photon parity measurements and on/off measurements.
We show that the squeezing operation can increase the degrees of Bell violations when
photon parity measurements are used, while
it depends on the values of amplitudes and squeezing for the case of
photon on/off measurements.
We also point out that
that fidelity of the generated SCS should be improved
up to at least 92\% with respect to ideal state in order to demonstrate direct Bell violations in real 
experiments.

This paper is organized as follows.
In Sec. \textrm{II}, two different approaches to entangle and
squeeze SCSs are briefly presented.
One is to pass a squeezed SCS through a beam splitter
to generate an entangled state,
and the other is to apply the two-mode squeezing operation on an ECS.
We then analyze, in Sec. \textrm{III}, the effects of the single-mode and
two-mode squeezing for Bell inequality tests.
In Sec. \textrm{IV}, we apply our theoretical evaluation to experimentally
feasible squeezed SCSs considering experimental imperfections.
A summary is given in Sec. \textrm{V} with final remarks on prospects
for experimental tests of Bell inequalities using SCSs.

\setlength\arraycolsep{1pt}

\section{Entangling and squeezing superpositions of coherent states}

\begin{figure}[t]
\centerline{\scalebox{0.35}{\includegraphics{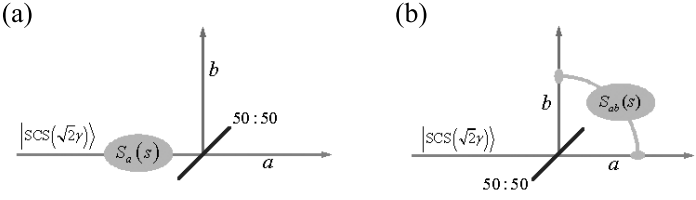}}}
\caption{Entangling and squeezing procedure for (a) an ESS and (b)
a SECS. The ESS is obtained by single-mode-squeezing $\ket{\mathrm{SCS}_{\pm}
  \left(\! {\scriptstyle \sqrt{2}}\gamma \right)}$ and feeding it into
a 50:50 beam splitter, whereas the SECS by feeding $\ket{\mathrm{SCS}_{\pm}
   \left(\! {\scriptstyle \sqrt{2}}\gamma \right)}$ into a 50:50 beam splitter and two-mode-squeezing 
it.}
\label{Squeezing}
\end{figure}

We introduce two particular types of SCSs, namely, even and odd SCSs, as
\begin{equation}
\ket{\mathrm{SCS}_{\pm} \left( \gamma \right)}  = \mathcal{N}_{\pm} \!\left( \gamma \right)
\left( \ket{\gamma} \pm \ket{-\gamma} \right)
\label{EO}
\end{equation}
where
$\mathcal{N}_{\pm}$
are normalization factors, $\ket{\gamma}$ is a coherent state of amplitude $\gamma$, and
$\gamma$ is assumed to be real for simplicity without loss of generality.
The SCS with the plus (minus) sign between the coherent states in Eq.~(\ref{EO}) is called
an even (odd) SCS because it contains an even (odd) number of photons regardless
of the value of $\gamma$.
The size of a SCS may be defined by the magnitude of the amplitude $\gamma$.
The ECSs at modes $a$ and $b$ are defined as
\begin{eqnarray}
\ket{\Phi_\pm} &=& \mathcal{N}_{\pm} \left( \ket{\gamma}_a \ket{\gamma}_b  \pm \ket{-\gamma}_a 
\ket{-\gamma} _b  \right),
\nonumber \\
\ket{\Psi_\pm} &=& \mathcal{N}_{\pm}  \left( \ket{\gamma}_a \ket{-\gamma}_b \pm \ket{-\gamma}_a 
\ket{\gamma} _b  \right),
\end{eqnarray}
which can be generated by splitting $\ket{\mathrm{SCS}_{\pm} \left( {\scriptstyle \sqrt{2}} \gamma 
\right)}$
at a 50:50 beam splitter with an appropriate phase.
We refer to the normalization factor $\mathcal{N}_{\pm}$ as $\mathcal{N}_{\pm} ( {\scriptstyle \sqrt{2}} 
\gamma )$ hereafter.
Note that $\ket{\Phi_{-}}$ and $\ket{\Psi_{-}}$ are maximally entangled ({\it i.e.}, each of them 
contains 1 ebit),
which in general show stronger Bell violations than $\ket{\Phi_{+}}$ and $\ket{\Psi_{+}}$ \cite{Jeong01}.

For Bell inequality tests, we shall use two types of entangled states, {\it i.e.}, entangled squeezed 
SCSs
(ESSs) and squeezed ECSs (SECSs). The former can be obtained by beam-splitting after 
single-mode-squeezing SCSs,
and the latter by two-mode-squeezing after beam-splitting a SCSs as shown in Fig.~\ref{Squeezing}.
The squeezed SCSs (SSCS) and ESSs can be represented as
\begin{align}
\ket{\mathrm{SSCS }_{\pm}\left( \gamma \right)} = S \left( s \right) \ket{\mathrm{SCS}_{\pm}  \left( 
\gamma \right))}, \\
\ket{\psi_{\pm}} = B_{ab} ~
\ket{\mathrm{SSCS}_{\pm} \left(\! {\scriptstyle \sqrt{2}}\gamma \right)}_a \ket{0}_b,
\label{eq4}
\end{align}
where $S_a \! \left( s \right) \!=\! \exp \!\left[ \frac{s}{2} \left( a^2 - a^{\dag 2}\right) \right]$
is the single mode squeezing operator,
$ B_{ab} \!=\! \exp \!\left[ \frac{\pi}{4} \left(a^\dag b - a^\dag b \right) \right] $  the 50:50 beam 
splitter operator, and $a$ and $a^\dagger$ ($b$ and $b^\dagger$) the bosonic annihilation
and creation operators for mode $a$ (mode $b$). The ESSs become the same as $\ket{\Psi_\pm}$
for the case of $s=0$. The SECSs are
\begin{eqnarray}
\ket{\Phi^s_{\pm}} &=& S_{ab} \! \left( s \right) \ket{\Phi_{\pm}}, \nonumber \\
\ket{\Psi^s_{\pm}} &=& S_{ab} \! \left( s \right) \ket{\Psi_{\pm}},
\end{eqnarray}
where $S_{ab} \! \left( s \right) \!=\! \exp \!\left[ s \left(a b - a^\dag b^\dag \right) \right]$
is the  two-mode squeezing operator. We assume that the squeezing parameter $s$ is real for
both $S_a(s)$ and $S_{ab}(s)$. The corresponding state is then squeezed along the real axis in
the phase space for $s>0$ while it is squeezed along the imaginary axis  for $s<0$.

\section{Violations of Bell's Inequality with photon parity and on/off measurement schemes}

\subsection{Bell-CHSH inequality with the Wigner functions}

Banaszek and W\'{o}dkiewicz (BW) studied Bell's inequality in the phase space, in terms of
the Wigner ($Q$) function based upon photon number parity  (on/off) measurements and the
displacement operation \cite{BW}. The Wigner function approach is based upon Clauser, Horne,
Shimony and Holt (CHSH)'s version of Bell's inequality while the $Q$ function upon Clauser
and Horne (CH)'s \cite{BW}. The displaced parity operator used for the Bell-CHSH inequality is
\begin{align}
\mathcal{P} (\alpha )  &=  \Pi^{\mathrm{even}} \! \left(\alpha \right) -  \Pi^{\mathrm{odd}}\! 
\left(\alpha \right)
\nonumber \\
&=   D (\alpha )\! \sum_{n=0}^\infty {\left( \ket{2n} \! \bra{2n} -\ket{2n\!+\!1} \! \bra{2n\!+\!1} 
\right) }   D^\dagger  (\alpha ),
\end{align}
where $  D (\alpha ) = \exp ( \alpha a^\dag - \alpha^* {a} ) $
is the displacement operator, and the Bell operator is
\begin{eqnarray}
 \mathcal{B}_{\textrm{CHSH}} &=&  \mathcal{P} _{a} \left( \alpha \right) \otimes \mathcal{P} _{b}
  \left( \beta \right)  +  \mathcal{P} _{a} \left( \alpha' \right)\otimes  \mathcal{P} _{b}   \left( 
\beta \right)
\nonumber \\
& & + ~ \mathcal{P} _{a} \left( \alpha \right) \otimes \mathcal{P} _{b} \left( \beta' \right)
-  \mathcal{P} _{a} \left( \alpha' \right) \otimes \mathcal{P} _{b} \left( \beta' \right).
\end{eqnarray}
The Wigner functions for state $\rho$ may be obtained by taking the average of the parity operator
$\mathcal{P} \left(\alpha \right)$ as \cite{Barnett,BW}
\begin{equation}
W \left(\alpha \right) = \frac{2}{\pi} ~\mathrm{Tr}
\left[ \,\rho \mathcal{P} \left(\alpha \right) \right]
\label{wf}
\end{equation}
and for two-mode state $\rho_{ab}$ as
\begin{equation}
W \left(\alpha ,\beta \right) = \Big(\frac{2}{\pi}\Big)^2 ~\mathrm{Tr}
\left[ \,\rho_{ab} \mathcal{P}_a \left(\alpha \right)\otimes
\mathcal{P}_b \left(\alpha \right) \right].
\label{wf2}
\end{equation}
Thus the Bell-CHSH inequality  can be represented by the Wigner function as
\begin{eqnarray} \label{BellInqualityWigner}
\left|B_{\mathrm{CHSH}} \right|  &=& \left(\frac{\pi}{2}\right)^2
| W \!\left(\alpha, \beta \right) + W\!\left(\alpha', \beta \right) +  W\!\left(\alpha, \beta' \right)
\nonumber \\
& & - W \!\left(\alpha', \beta' \right) | \leq 2,
\end{eqnarray}
where $ W\!\left(\alpha, \beta \right)$ is the two-mode Wigner function
and we refer to $B_{\textrm{CHSH}} \!=\! \langle \mathcal{B}_{\textrm{CHSH}} \rangle$ as the Bell-CHSH 
function.
This inequality can be violated with appropriate measurement operators and entangled states, and its
maximum value, $2\sqrt{2}$, is known as Cirel'son's bound \cite{Cirel'son}.

Using Eqs.~(\ref{EO}) and (\ref{wf}), the Wigner functions for the even and odd SCSs can be calculated as
\begin{equation}
W^{\mathrm{SCS}}_{\pm} \!\left(\alpha \right) = \mathcal{N}_{\pm}^{\,2} \left[
W_{\!{\scriptstyle \sqrt{2}} \gamma} \!\left(\alpha \right) + W_{\!{\scriptstyle -\sqrt{2}} \gamma} 
\!\left(\alpha \right)
\pm 2 X_{\!{\scriptstyle \sqrt{2}} \gamma} \!\left(\alpha \right) \right],
\label{eq9}
\end{equation}
where $W_{\!\gamma} \left(\alpha \right) =2 e^{ -2|\alpha -\gamma|^2 }/\pi$
 is the Wigner function of coherent state $\ket{\gamma}$
and
$X_{\!\gamma} \left(\alpha \right)
=2 e^{ -2|\alpha|^2 } \cos \left[ 4 \, \mathrm{Im}\, (\alpha^* \gamma) \right]/\pi$.
The Wigner functions of the ESSs can be obtained using Eqs.~(\ref{eq4}) and (\ref{wf2}), and they
can also be expressed as
\begin{equation}
W_{\psi_{\pm}} \!\left(\alpha, \beta \right) =   W^{\mathrm{SCS}}_{\pm}
\!\left( \frac{\alpha^s-\beta^s}{\sqrt 2} \right) W_{0}
\!\left(\frac{\alpha+\beta}{\sqrt 2} \right),
\end{equation}
where
$W_0(\alpha)$ is the Wigner function of the vacuum and
the superscript $s$ is used to indicate
\begin{equation}
\alpha^s = \alpha \cosh s + \alpha^* \sinh s
= e^s \mathrm{Re} \,\alpha + i \, e^{-s} \mathrm{Im} \,\alpha.
\end{equation}
for an arbitrary complex number $\alpha$.
The two-mode Wigner functions for the ECSs are calculated in the same manner as \cite{Jeong01}
\begin{eqnarray}
W_{\Phi_{\pm}}  (\alpha, \beta) &=&   \mathcal{N}_{\pm}^{\,2} \Big[
W_{\gamma}  (\alpha) W_{\gamma}  (\beta) +
W_{-\gamma} (\alpha) W_{-\gamma} (\beta)
\nonumber \\
& \pm &  2 X_{\gamma} (\alpha) X_{\gamma} (\beta)
\mp 2 Y_{\gamma} (\alpha) Y_{\gamma} (\beta) \Big],
\nonumber \\
W_{\Psi_{\pm}} (\alpha, \beta) &=&   \mathcal{N}_{\pm}^{\,2} \Big[
W_{\gamma} (\alpha) W_{-\gamma} (\beta) +
W_{-\gamma} (\alpha) W_{\gamma} (\beta)
\nonumber \\
& \pm &  2 X_{\gamma} (\alpha) X_{\gamma} (\beta)
\pm 2 Y_{\gamma} (\alpha) Y_{\gamma} (\beta) \Big],
\label{eq12}
\end{eqnarray}
where $Y_{\gamma} \left(\alpha \right)
= 2 e^{ -2|\alpha|^2 } \sin \left[ 4 \, \mathrm{Im}\, (\alpha^* \gamma) \right] / \pi$.
The Wigner functions for SECSs are then
\begin{eqnarray}
W_{\Phi^s_{\pm}} (\alpha, \beta) &=&  W_{\Phi_{\pm}}  (\tilde{\alpha}^s, \tilde{\beta}^s),
\nonumber \\
W_{\Psi^s_{\pm}} (\alpha, \beta) &=&  W_{\Psi_{\pm}}  (\tilde{\alpha}^s, \tilde{\beta}^s),
\end{eqnarray}
where
\begin{equation}
\tilde{\alpha}^s \!=\! \alpha \cosh s+ \beta^* \sinh s, ~ \tilde{\beta}^s =  \beta \cosh s + \alpha^* 
\sinh s .
\end{equation}
Note that when $s=0$, $W_{\Psi^s_{\pm}} (\alpha, \beta)=W_{\psi_{\pm}} (\alpha, \beta)$ and 
$W_{\Phi^s_{\pm}} (\alpha, \beta)=W_{\psi_{\pm}} (-\beta, \alpha)$.

\begin{figure}[t]
\centerline{\scalebox{0.365}{\includegraphics{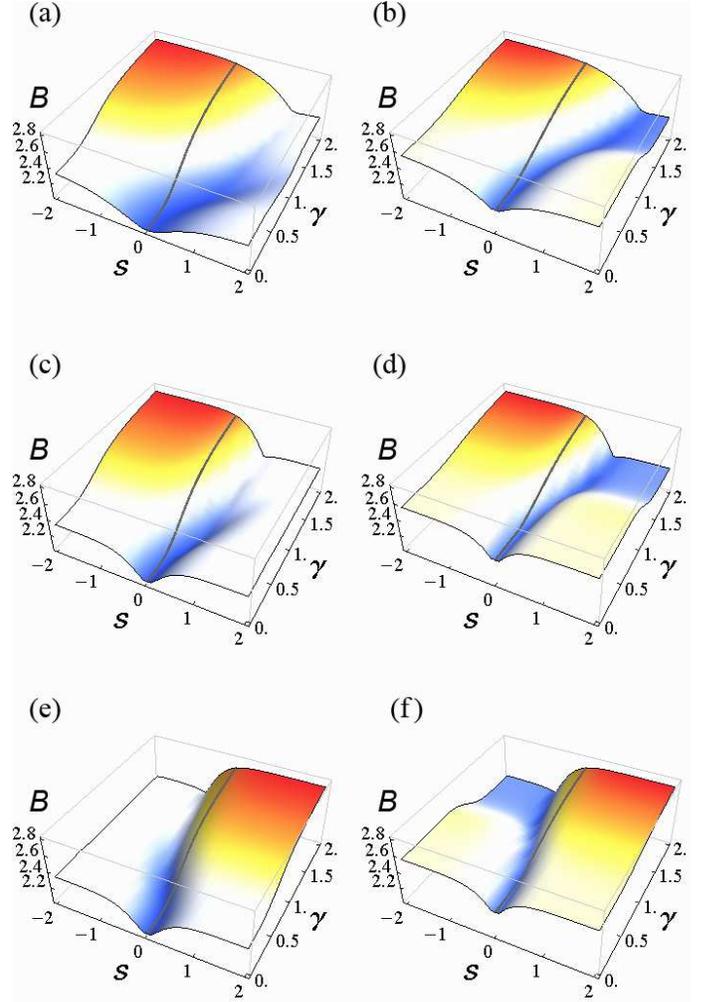}}}
\caption{(Color online) Optimized Bell-CHSH function $B=|B_{\mathrm{CHSH}}|_{\mathrm{max}}$ for (a) 
$\psi_+$ (b) $\psi_-$ (c) $\Phi^s_+$ (d) $\Phi^s_-$ (e) $\Psi^s_+$ (f) $\Psi^s_-$ for parity 
measurements.
The split line in each graph indicates no squeezing ($s=0$). Note that the plots for $\psi_{\pm}$ are 
similar to the ones for $\Phi^s_{\pm}$, and that $B$'s of $\Psi^s_{\pm}$ are rather similar and symmetric 
to the ones $\Phi^s_{\pm}$ with respect to $s=0$ line.
One can observe that for small $\gamma$ squeezing in any direction can enhance Bell violations, whereas 
for large $\gamma$ squeezing in specific direction only can enhance them. In any case, squeezing causes 
Bell violations to increase monotonically from the non-squeezed values and converge to specific ones.}
\label{BmaxParity}
\end{figure}

It is known that the Bell violation for an ECS approaches Cirel'son's bound \cite{Cirel'son}
when the amplitude $\gamma$ becomes large \cite{Jeong01}.
Figure~\ref{BmaxParity} shows that a couple of characteristic properties in common
when squeezing is applied to the states being considered.
The squeezing operation increases the degree of the Bell violation up to some extent for small $\gamma$,
but has a tendency of degrading it for squeezing in the specific direction for larger $\gamma$.
For example, the squeezing in both the real and imaginary directions in the phase space, enhances 
violation for $\psi_{+}$ ($\psi_{-}$) until $\gamma$ reaches around 0.5 (1.0). On the other hand, for 
larger values of $\gamma$ squeezing in the real direction ({\it i.e.}, $s>0$) decreases the degree of 
violation while squeezing in the imaginary direction ({\it i.e.}, $s<0$) increases the violation.

In fact, for larger $\gamma$, squeezing along the real axis makes the interference fringes less sharp and 
this
could be related to the decrease of the Bell violations.
In the case of $\Phi^s_{\pm}$ with large $\gamma$, since $\tilde{\alpha}^s (\tilde{\beta}^s) \rightarrow 
\alpha ' (-\alpha '^*)$ as $s\rightarrow-\infty$ where $\alpha' = \frac{1}{2} e^{-s} (\alpha-\beta^*)$, 
and hence $W_{\Phi^s_{\pm}} (\alpha, \beta) \rightarrow W_{\Phi_{\pm}}  (\alpha', -\alpha '^*)$ which is 
the very condition when maximum violations occur for $W_{\Phi_{\pm}}  (\alpha, \beta)$.
But as $s\rightarrow \infty$, the interference part $X_{\gamma} (\tilde{\alpha}^s) X_{\gamma} 
(\tilde{\beta}^s) - Y_{\gamma} (\tilde{\alpha}^s) Y_{\gamma} (\tilde{\beta}^s)$ in the Wigner function 
fades out, which may play a crucial role in degrading the Bell violations.
The case of $\Psi^s_{\pm}$ can be explained in a similar way.
Therefore, in the case of photon parity measurements, squeezing in a well-chosen quadrature direction can 
enhance Bell violations of tested states, though its contribution gets slighter as the amplitudes of the 
states grow larger.

\subsection{Bell-CH inequality with the $Q$ functions}

The operator used for tests of the Bell-CH inequality  is
\begin{eqnarray}
 \mathcal{B}_{\textrm{CH}} &=&  \mathcal{Q} _{a} \left( \alpha \right) \otimes \mathcal{Q} _{b} \left( 
\beta \right) +  \mathcal{Q} _{a} \left( \alpha' \right) \otimes \mathcal{Q} _{b} \left( \beta \right)
\nonumber \\
& & + \, \mathcal{Q} _{a} \left( \alpha \right) \otimes \mathcal{Q}_{b} \left( \beta' \right)
- \mathcal{Q}_{a} \left( \alpha' \right) \otimes \mathcal{Q} _{b} \left( \beta' \right)
\nonumber \\
& &
- \, \mathcal{Q}_{a} \left( \alpha \right) \, \otimes \mathcal{I}_b - \mathcal{I}_a \,\otimes 
\mathcal{Q}_{b} \left( \beta \right),
\end{eqnarray}
where
\begin{equation}
\mathcal{Q}(\alpha ) =  D (\alpha ) \ket{0} \!\bra{0}  D^\dag (\alpha )
\end{equation}
is a displaced ``no photon" operator and $\mathcal{I}$ is the identity operator.
Subsequently, the Bell-CH function $B_{\mathrm{CH}} = \langle \mathcal{B}_{\textrm{CH}} \rangle$ is given 
in terms of $Q$ representation as
\begin{eqnarray}
B_{\mathrm{CH}} &=& \pi^2 \big[ Q_{ab} \left(\alpha, \beta \right) + Q_{ab} \left(\alpha', \beta \right) 
+ Q_{ab} \left(\alpha, \beta' \right)
\nonumber \\
& & - \, Q_{ab} \left(\alpha', \beta' \right) \!\big] - \pi \big[ Q_a \left(\alpha \right) + Q_b 
\left(\beta \right) \!\big],
\end{eqnarray}
where  $Q_a \left(\alpha \right)$ and $Q_b \left(\beta \right)$ are marginal $Q$ functions in the 
corresponding modes.
As implied above, the $Q$ functions of single-mode state $\rho$
and two-mode state $\rho_{ab}$
can be obtained using the operator $\mathcal{Q}(\alpha )$
as $(1/\pi){\rm Tr}[\rho \mathcal{Q}(\alpha )]$
and $(1/\pi)^2{\rm Tr}[\rho_{ab} \mathcal{Q}_a (\alpha )\otimes\mathcal{Q}_b (\beta )]$, respectively
\cite{Barnett}.
The $Q$ functions for the SSCS are then given as
\begin{equation}
Q_{\pm}^{\mathrm{SSCS}} \!\left(\alpha \right) = \, \mathcal{N}_{\pm}^{\,2} \big[ Q^{+}_{\!\sqrt{2} 
\gamma} \!\left(\alpha \right) + Q^{-}_{\!\sqrt{2} \gamma} \!\left(\alpha \right)
\pm 2 \, Q^{\mathrm{X}}_{\!\sqrt{2} \gamma} \!\left(\alpha \right)  \big],
\label{QS}
\end{equation}
subsequently for ESSs as
\begin{equation}
Q_{\psi_{\pm}} \!\left(\alpha, \beta \right) = Q^{\mathrm{SSCS}}_{\pm}
\!\left( \frac{\alpha-\beta}{\sqrt 2} \right) Q_{0} \!\left(\frac{\alpha+\beta}{\sqrt 2} \right),
\end{equation}
where
\begin{align}
Q^{\pm}_{\gamma} \left(\alpha \right)
&= \cos \theta ~ Q_{\pm \gamma_{-s} } (\alpha_{s} ),
\\
Q^{\mathrm{X}}_{\gamma} \left(\alpha \right)
&= \cos \theta ~ Q_{0} (\alpha_{s} ) e^{-|\gamma_{-s}|^2}
\cos \left[ 2 \,\mathrm{Im} (\alpha^*_s \gamma_s) \right],
\end{align}
where
$\alpha_{s} = \alpha \cos (\theta/2) + \alpha^* \sin (\theta/2)$,
$\gamma_{-s} = \gamma \cos (\theta/2) - \gamma^* \sin (\theta/2)$,
$\theta/2 = \tan^{-1} \left( \tanh \frac{s}{2} \right)$,
$Q_{\gamma} (\alpha) =  e^{-|\alpha-\gamma|^2}/{\pi}$, and $Q_{0} (\alpha) =  e^{-|\alpha|^2}/{\pi}$.
Note that as $s \rightarrow \infty \, (-\infty)$, $\alpha_{s} \rightarrow \sqrt{2}
\, \mathrm{Re}\, [\alpha] ~\left(i \sqrt{2}\, \mathrm{Im} \, [\alpha] \right)$.

In the meantime, the $Q$ functions for SECSs are
\begin{eqnarray}
Q_{\Phi^s_{\pm}} \left(\alpha, \beta \right) &=& \mathcal{N}_{\pm}^{\,2} \big[ Q_{++} \left(\alpha, \beta 
\right) + Q_{--} \left(\alpha, \beta \right)
\nonumber \\
& & \pm \, 2\, Q^{\mathrm{XY}}_{+} \left(\alpha, \beta \right) \big],
\label{QQ1}\\
Q_{\Psi^s_{\pm}} \left(\alpha, \beta \right) &=& \mathcal{N}_{\pm}^{\,2} \big[Q_{+-} \left(\alpha, \beta 
\right) + Q_{-+} \left(\alpha, \beta \right)
\nonumber \\
& & \pm \, 2\, Q^{\mathrm{XY}}_{-} \left(\alpha, \beta \right)  \big] ,
\label{QQ2}
\end{eqnarray}
where
\begin{eqnarray}
Q_{\cal \pm\pm} \left(\alpha, \beta \right)
&=& {\cos}^2 \theta ~
Q_{\pm \gamma_{\mp s}} (\tilde{\alpha}_{s} ) Q_{\pm \gamma_{\mp s}} (\tilde{\beta}_{s} ),
\nonumber \\
Q_{\cal \pm\mp} \left(\alpha, \beta \right)
&=& {\cos}^2 \theta ~
Q_{\pm \gamma_{\mp s}} (\tilde{\alpha}_{s} ) Q_{\mp \gamma_{\pm s}} (\tilde{\beta}_{s} ),
\\
Q^{\mathrm{XY}}_{\pm} \left(\alpha, \beta \right)
&=& {\cos}^2 \theta \, Q_{0} (\tilde{\alpha}_{s} ) Q_{0} (\tilde{\beta}_{s} ) e^{-2|\gamma_{\mp s}|^2}
\nonumber \\ & &\times
 \cos \left[ 2 \,\mathrm{Im} ( \alpha^*_s \gamma_s \pm \beta^*_s \gamma_s ) \right],
\end{eqnarray}
with $\tilde{\alpha}_{s} = \alpha \cos (\theta/2) + \beta^* \sin (\theta/2)$ and
$\tilde{\beta}_{s} =  \beta \cos(\theta/2)  + \alpha^* \sin (\theta/2)$.
Since the Bell-CH inequality is equivalent to the Bell-CHSH inequality
for the case of bipartite systems and dichotomic measurements, the above Bell-CH function can be replaced 
with the Bell-CHSH function, which shall be further clarified in the following subsection.

\subsection{Bell-CHSH inequality with on/off measurements}

One can test the Bell-CHSH inequality by the following displaced ``on/off\," measurement operator
\begin{eqnarray}
\mathcal{O}(\alpha ) &=&  \Pi^{\rm on} \! \left(\alpha \right) -  \Pi^{\rm off}\! \left(\alpha \right)
\nonumber \\
 &=&  D^\dag \! (\alpha )  \left( \sum_{n = 1}^\infty { \ket{n} \! \bra{n} }  - \ket{0} \! \bra{0} 
\right) \!  D (\alpha ),
\end{eqnarray}
which assigns +1 or --1 to each measured result depending on whether (any) photons are detected or not
at a detector such as an avalanche photodiode.
Then the Bell-CHSH inequality can be represented in the same way as done in 
Eq.~(\ref{BellInqualityWigner}) just with $\mathcal{P}$ replaced with $\mathcal{O}$, so that the 
Bell-CHSH function becomes
\begin{equation}
B_{\mathrm{CHSH}} = A \!\left(\alpha, \beta \right) + A\!\left(\alpha', \beta \right) +  A\!\left(\alpha, 
\beta' \right) - A \!\left(\alpha', \beta' \right)
\end{equation}
with
\begin{eqnarray}
A \!\left(\alpha, \beta \right) &=& 1- 2 \pi\, Q_a \left(-\alpha \right) -2 \pi\, Q_b \left( -\beta 
\right)
\nonumber \\ & &
+ \, 4 \pi^2 \,Q_{ab} \left(-\alpha, -\beta \right),
\end{eqnarray}
where the $Q$ functions are the ones obtained in the previous subsection.
It is worth noting that this Bell-CHSH function is related to the previous Bell-CH function as
\begin{equation}
B_{\mathrm{CHSH}} \left(\alpha, \beta \right)= 4 B_{\mathrm{CH}} \left(-\alpha, -\beta \right) + 2 .
\end{equation}

\begin{figure}[t]
\centerline{\scalebox{0.365}{\includegraphics{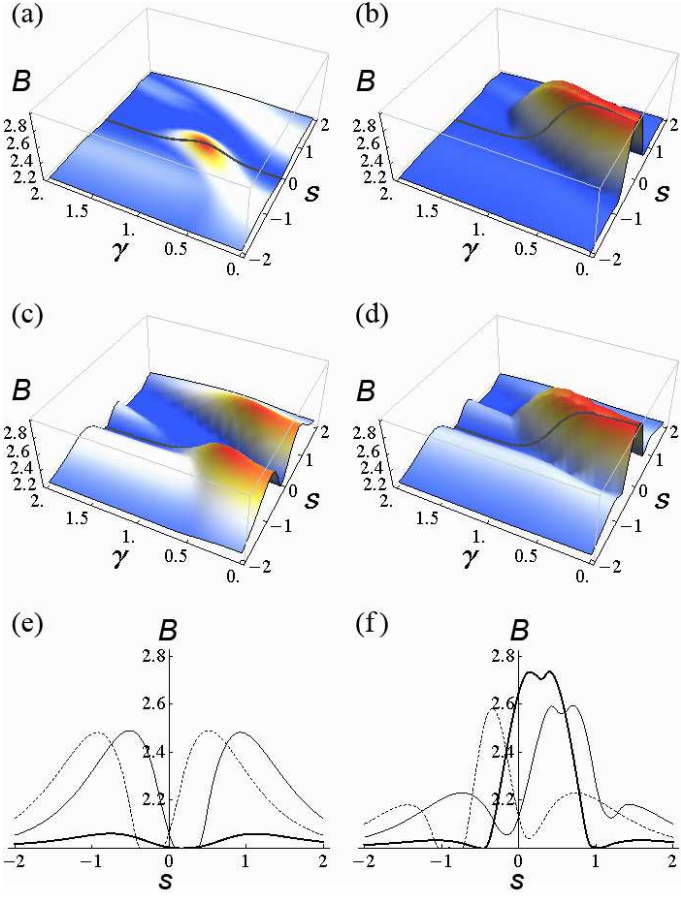}}}
\caption{(Color online) Optimized Bell-CHSH function $B=|B_{\mathrm{CHSH}}|_{\mathrm{max}}$ for (a) 
$\psi_+$ (b)
$\psi_-$ (c) $\Phi^s_+$ (d) $\Phi^s_-$ for on/off measurements. The lowest two plots are for (e)
$\psi_+$ (thick), $\Phi^s_+$ (thin), $\Psi^s_+$ (dashed) respectively with $\gamma=0.5$ and for
(f) $\psi_-$ (thick), $\Phi^s_-$ (thin), $\Psi^s_-$ (dashed) with $\gamma=1.0$. The
plots of $\Psi^s_{\pm}$ not presented here are similar to those of $\Phi^s_{\pm}$
provided the sign of $s$ is altered as in the parity measurement case. Note that
for small $\gamma$, squeezing ``+" states increases $B$ up to some extent (e), and that for
large $\gamma$, squeezing ``--" states in specific direction only contributes to maximal
values of $B_{\mathrm{CHSH}}$'s (f).}
\label{BmaxOnOff}
\end{figure}

A Bell inequality test with photon on/off measurements is obviously more feasible than
that of photon number parity measurements. However, if the average photon number
of the state under consideration is too large,
Bell violations cannot be observed using photon on/off measurements because
the possibility of getting a ``off" result approaches zero \cite{Jeong01}.
Because of this, Bell violations for ESSs and SECSs shown in Fig. \ref{BmaxOnOff}
show different behaviors compared to the cases of photon parity measurements.
In the case of ``+" states ($\Phi_{+},~ \Psi_{+} (\psi_{+})$), quadrature squeezing in any direction 
increases Bell violations only for small $\gamma$. Meanwhile, in the case of ``--'' states ($\Phi_{-},~ 
\Psi_{-} (\psi_{-})$), squeezing in specific direction increases the violations only for $\gamma \gtrsim 
1$, whereas it is not any desirable for violations for small $\gamma$.
In any case, large squeezing in any quadrature direction causes Bell violations to eventually vanish. 
This is different from the cases for the parity measurements where large values of squeezing cause the 
Bell functions to converge to certain values (smaller or larger than the ones in the cases of no 
squeezing).

\begin{figure}[t]
\centerline{\scalebox{0.3}{\includegraphics{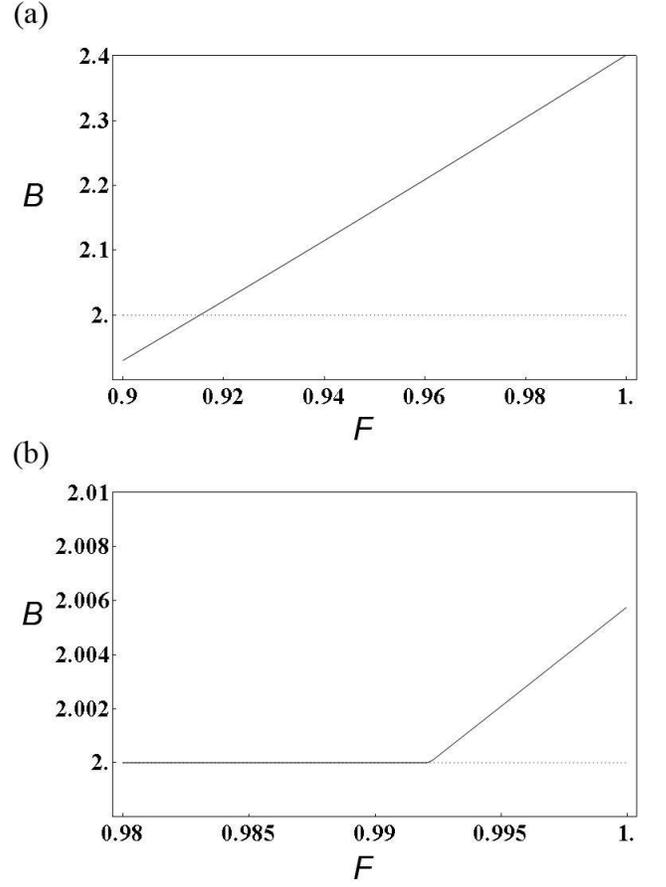}}}
\caption{Optimized Bell-CHSH function $B=|B_{\mathrm{CHSH}}|_{\mathrm{max}}$ vs. fidelity $F$ of 
$\rho_\mathrm{exp}$ with respect to $\ket{\phi_2}$ for the case of (a) photon parity measurements and (b) 
photon on/off measurements. Dotted line in each plot indicates the local realistic bound for Bell-CHSH 
inequality.
 Bell violation for the case of parity measurements can be observed when the fidelity approaches 92\% 
while  that for on/off measurements case cannot be observed until the fidelity goes over 99\%.}
\label{BmaxExp}
\end{figure}

\section{Estimation of Bell violations with realistic states}

We are also interested in whether a recently generated SSCS \cite{Ourjoumtsev},
which can be immediately used to generate an ESS, may be used for tests of
Bell's inequality.
The size of the generated SSCS, an ``even" one, was as large as $\gamma=\sqrt{2.6}$
and the squeezing degree was 3.5dB along the real axis in the phase space.
The ideal state that can be generated with a two-photon number state using the scheme described in
Ref. \cite{Ourjoumtsev} is \cite{OurjoumtsevCorrected}
\begin{equation}
\ket{\phi_2} = \sqrt{2/3} \ket{2} + \sqrt{1/3} \ket{0}.
\label{our}
\end{equation}
This state is a very good approximation of an ideal SSCS
\begin{equation}
\ket{\mathrm{SSCS}_{+}} = S \left( s_0  \right)
\ket{\mathrm{SCS}_{+} \left( \alpha_0 \,\right)}
\label{ideal}
\end{equation}
where $s_0=0.4$, $\alpha_0=\sqrt{2.6}$ and
the fidelity between the two states is as high as $\left| \langle \phi_2 \ket{\mathrm{SSCS}_{+}} 
\right|^2 \approx 99$\%.
If state (\ref{ideal}) is injected into a 50:50 beam splitter, it becomes $\ket{\psi_+}$ with 
$\gamma=\sqrt{2.6/2}$.
When this ideal two-mode state is used to obtain the Bell function $B_{\mathrm{CHSH}}$, its optimized 
value is 2.419 (2.033) with photon number parity (on/off) measurements.
In the meantime, state $\ket{\phi_2}$ shows a Bell inequality violation as large as 
$B_{\mathrm{CHSH}}=2.401$ ($2.006$) using parity (on/off) measurements.
In order to analyze the case of the actually generated (mixed) state $\rho_{\mathrm{exp}}$, which is 
degraded by experimental imperfections such as non-unit efficiencies, noises, and errors related to 
measuring devices, we use the following Wigner function in Ref. \cite{Ourjoumtsev},
\begin{widetext}
\begin{align}
W_{\mathrm{exp}} \left( x,\, p \right) ~=~& \frac{\exp \left( - x^2 /\alpha  - p^2 /\beta \right)}{\pi 
\sqrt {\alpha \beta } \left[ \left( 1 - \frac{\delta \alpha \left( 1 - \nu \right)^2 }{2\left( {\alpha  - 
\beta } \right)} \right)^2  + \frac{1}{2}\left( \frac{\delta \alpha \left( 1 - \nu \right)^2 }{2\left( 
{\alpha  - \beta } \right)} \right)^2 \right]}
\Bigg\{ \frac{\delta^2}{2}\left[ \frac{x^2}{\alpha} + \frac{\alpha \nu^2 p^2}{\beta^2} \right]^2
\nonumber \\
& + 2\delta \left[ 1 - \delta \left( 1 + \frac{\left( \alpha \nu  - \beta \right)^2}{2 \beta \left( 
\alpha  - \beta  \right)} \right) \right] \left[ \frac{x^2}{\alpha} + \frac{\alpha \nu ^2 p^2}{\beta ^2} 
\right] + \delta ^2 \frac{\left( \alpha \nu  - \beta  \right)^2}{2\beta \left( \alpha  - \beta  
\right)}\left[ \frac{x^2}{\alpha} - \frac{\alpha \nu ^2 p^2}{\beta ^2} \right]
\nonumber \\
& + \left[ 1 - \delta \left( 1 + \frac{\left( \alpha \nu  - \beta  \right)^2}{2\beta \left( \alpha  - 
\beta \right)} \right) \right]^2  + \frac{\delta ^2 }{2}\left[ \frac{\left( \alpha \nu  - \beta \right)^2 
}{2\beta \left( \alpha  - \beta  \right)} \right]^2   \Bigg\},
\label{WignerO}
\end{align}
\end{widetext}

where $x=\sqrt{2} \,\mathrm{Re}\, [\alpha], ~p=\sqrt{2} \,\mathrm{Im}\, [\alpha]$ in our case, and the 
four parameters $\alpha,~\beta,~\nu,~\delta$ are defined by gain and various imperfection parameters 
\cite{typos}.
However, since the Bell function depends very sensitively on such imperfection parameters, we assume
perfect measuring devices with no errors, in which case
\begin{equation}
\alpha \rightarrow g,~ \beta \rightarrow \alpha - (g-1)^2/g,~ \nu \rightarrow 1/g, ~\delta \rightarrow 1,
\end{equation}
where $g$ is an optical parametric amplifier gain describing the two-photon number state,
and then the fidelity $F=\bra{\phi_2} \rho_{\mathrm{exp}} \ket{\phi_2}$ depends only on $g$.
Note also that for testing the on/off measurement case, we can transform the above Wigner function into 
the $Q$ function simply by just replacing the parameters $\alpha,\, \beta,\, \delta$ by $\alpha+1,\, 
\beta+1,\, \frac{\alpha}{\alpha+1} \delta$.
As can be seen in Fig. \ref{BmaxExp}, in order for $\rho_{\mathrm{exp}}$ to show Bell violations,
the fidelity should be improved up to around 92\%
in the case of parity measurements. However, we note again that the violations are possible
only when all the experimental imperfections nearly vanish, which
is extremely demanding.
When on/off measurements are used, the fidelity should be even more
improved up to at least 99\% to show a Bell violation.

\section{Remarks}

We have studied how squeezing influences the degree of Bell inequality violations
of several different beam-split-entangled SCSs.
It has been found that squeezing can always increase Bell violations,
given the squeezing direction is properly chosen, for the case of photon parity measurements.
On the other hand, in the case of the photon on/off measurements, squeezing can enhance
Bell violations only for well-chosen values of amplitudes and squeezing.
Therefore, it should be noted that for certain measurement schemes, the squeezing action is not always 
helpful in enhancing Bell violations of entangled states of light.

In order to demonstrate a Bell violation in a real experiment, a significant
improvement is required over the currently available SSCS.
For example, the fidelity of the generated
state should be improved up to 92\% even when all the other conditions including
the efficiency of photon parity measurements are ideal.
There are ongoing efforts to effectively generate high-fidelity SSCSs
using currently available experimental resources \cite{mjk08}.
It would be a more realistic target to perform homodyne tomography to reconstruct
a generated SSCS and ``indirectly'' show a Bell violation using Eq. (7).

\acknowledgments

This work was supported by the World Class University (WCU) program and the
KOSEF grant funded by the Korea government(MEST) (R11-2008-095-01000-0).


\begin{thebibliography}{9}
\bibitem{OurjoumtsevCorrected} $\ket{\phi_2} = \sqrt{2/3} \ket{2} - \sqrt{1/3} \ket{0}$
in the original paper should be corrected to be the one we give here.
\bibitem{typos} Some minor typos in Ref. \cite{Ourjoumtsev} have been corrected in the Wigner function
(\ref{WignerO}).
\end{thebibliography}
\end{document}